\documentclass{Interspeech}

\interspeechcameraready

\title{Multimodal Biomarkers for Schizophrenia: Towards Individual Symptom Severity Estimation}

\author[affiliation={1}]{Gowtham}{Premananth}
\author[affiliation={2}]{Philip}{Resnik}
\author[affiliation={3}]{Sonia}{Bansal}
\author[affiliation={3}]{Deanna}{L.Kelly}
\author[affiliation={1}]{Carol}{Espy-Wilson}

\affiliation{Department of Electrical \& Computer Engineering}{University of Maryland, College Park}{USA}
\affiliation{Institute for Advanced Computer Studies}{University of Maryland, College Park}{USA}
\affiliation{School of Medicine}{University of Maryland}{USA}
\email{gowtham8@umd.edu, resnik@umd.edu,sbansal@som.umaryland.edu, dlkelly@som.umaryland.edu, espy@umd.edu}

\usepackage{comment}
\usepackage{multicol}
\usepackage{multirow}
\usepackage{diagbox}

\begin{document}

\maketitle

\begin{abstract}

Studies on schizophrenia assessments using deep learning typically treat it as a classification task to detect the presence or absence of the disorder, oversimplifying the condition and reducing its clinical applicability. This traditional approach overlooks the complexity of schizophrenia, limiting its practical value in healthcare settings. This study shifts the focus to individual symptom severity estimation using a multimodal approach that integrates speech, video, and text inputs. We develop unimodal models for each modality and a multimodal framework to improve accuracy and robustness. By capturing a more detailed symptom profile, this approach can help in enhancing diagnostic precision and supports personalized treatment, offering a scalable and objective tool for mental health assessment.
  
\end{abstract}

\section{Introduction}

Schizophrenia is a complex and heterogeneous mental health disorder that affects approximately 24 million people worldwide \cite{institute2021global}. Despite its significant impact, many individuals with schizophrenia and other mental health conditions remain undiagnosed or experience delays in receiving appropriate treatment. Limited access to mental healthcare, particularly in under-served and rural areas, exacerbates this issue, as mental health specialists and resources are often scarce \cite{saxena2007resources}. Even when care is accessible, the diverse and overlapping symptoms of schizophrenia pose a diagnostic challenge, frequently leading to misdiagnosis or prolonged delays in treatment. Traditional diagnostic approaches rely heavily on subjective clinical assessments, which can vary between practitioners and lack consistency. These challenges underscore the urgent need for accessible, scalable, and objective measures that can aid in symptom assessment by identifying reliable biomarkers and clinical subtypes that can capture pathological homology missed by conventional diagnostic assessments. Advancements in this area would support early and accurate identification and characterization of schizophrenia symptom profiles, ultimately yielding meaningful neurobiological and behavioral targets for treatments, etiological investigations and patient outcomes. 

One promising avenue for objective mental health assessment lies in multimodal analysis, integrating speech, video,and text-based inputs\cite{seneviratne2022multimodal,embc,sch_gow}. Recent research has demonstrated that speech patterns, facial expressions, and linguistic features extracted from transcripts can serve as valuable, non-invasive indicators of mental health conditions, particularly in disorders such as major depressive disorder \cite{seneviratne21b_dep}, post-traumatic stress disorder \cite{kathan24_ptsd}, and schizophrenia \cite{premananth24_schizo}. Accordingly, these multimodal objective signals can be used in order to identify clinical subtypes that converge on symptom domains as assessed by more subjective clinical interviews. Speech-based methods analyze vocal patterns, acoustic features, and articulatory characteristics, while video analysis can capture facial expressions and gaze behavior indicative of underlying symptoms. Additionally, text from transcripts of natural speech or clinical interviews can provide insights into linguistic coherence, thought patterns, and semantic anomalies associated with schizophrenia. Integrating speech, video, and text-based modalities could improve the robustness and accuracy of symptom detection, allowing clinicians to tailor treatment strategies based on specific symptom severity. 

However, much of the existing research has framed these approaches as binary classification tasks—focusing on determining whether an individual has a disorder—rather than capturing the nuances of specific symptoms. Given symptomatology that schizophrenia is a highly heterogeneous disorder, shifting the focus toward estimating individual symptom profiles that are associated with objective through multimodal inputs could offer a more precise and clinically relevant approach, enabling more targeted interventions and personalized treatment strategies.

While some studies have explored overall severity estimation \cite{gowtham_spade}, this remains a relatively small subset of the field. A major challenge in multimodal assessment is that many mental health disorders share overlapping symptoms, making a simple classification or severity score insufficient for clinical decision-making. To enhance the practical utility of these models, the focus should shift toward individual symptom severity estimation, which provides a more granular understanding of a patient's condition. 

One critical consideration in developing such systems is whether to use multiple models for different symptoms or a single unified model capable of estimating all symptoms simultaneously. A unified model offers several advantages. First, schizophrenia symptoms often co-occur and influence one another, meaning that a model trained to recognize all symptoms collectively may better capture these inter-dependencies. Second, training and deploying separate models for each symptom would be computationally expensive and inefficient, requiring significant resources for both development and real-time inference. Third, a single model ensures greater consistency in symptom estimation, reducing variability that might arise when combining predictions from multiple independent models. Lastly, integrating all symptoms into a unified framework allows for a more holistic and clinically meaningful assessment, aligning more closely with real-world diagnostic practices where clinicians evaluate symptoms in conjunction rather than in isolation.

In this study we focus on doing an individual symptom severity estimation based on the symptoms exhibited by schizophrenia subjects. The main contributions of this study are:
\begin{enumerate}
     \item Unimodal models using speech,video, and text inputs to estimate individual schizophrenia symptoms.
     \item A multimodal framework involving speech,video, and text inputs to better perform individual schizophrenia symptoms estimation.
     \item We analyze the extent to which different modalities capture specific symptoms, providing insights into the strengths and limitations of each modality in symptom assessment.
\end{enumerate}

By adopting a unified model for symptom estimation, this study aims to improve the precision and efficiency of schizophrenia assessment. The findings could contribute to the development of more objective, scalable, and clinically useful diagnostic tools, ultimately aiding in earlier detection and better-informed treatment planning.

\section{Dataset}

The dataset used in this study is a dataset that was collected as a part of an interdisciplinary mental health assessment project jointly carried out by the University of Maryland School of Medicine and the University of Maryland - College Park \cite{Kelly2020-cj}. The dataset contains English audio and video recordings of subjects who were interviewed during their in-person clinic visits. The dataset contains data from a total of 140 sessions produced by 39 different subjects. Before their interview sessions they were also evaluated based on mental health assessment questionnaires like Hamilton Depression rating scale (HAMD)\cite{HAMD} and Brief Psychiatric Rating Scale (BPRS) \cite{overall1962brief} to quantify their symptom severity before their sessions. In this study we are focusing our experiments based on the 18-item BPRS scores as they look into the psychiatric symptoms presented by the subjects which are good indicators for schizophrenia. In the 18-item BPRS scale all of these symptoms are individually rated on a scale of 1-7 that goes up with increasing severity of the particular symptom. For example, a score of 1 stands for symptom not present, a score of 4 meaning moderate severity, and a score of 7 stands for extremely severe symptom. Table.\ref{tab:severity} gives a summary of the 18 symptoms and the frequency of the symptom scores across all the sessions in the dataset.

\begin{table}[h!]  
    \caption{\centering
    \textbf{BPRS symptoms and the frequency of the symptom scores in the dataset}}
    \label{tab:severity}
    \centering
    \begin{tabular}{lcccccc}
    \toprule
\diagbox[width=12em]{Symptom}{Severity score}&1   & 2  & 3  & 4  & 5  & 6  \\
\midrule
        Somatic Concern&45  & 44 & 39 & 10 & 2  & 0  \\
        Anxiety&25  & 39 & 52 & 20 & 2  & 2  \\
        Guilt&82  & 33 & 20 & 4  & 0  & 1  \\
        Grandiosity&82  & 24 & 4  & 11 & 11 & 8  \\
        Depression&67  & 35 & 22 & 14 & 2  & 0  \\
        Hostility&56  & 42 & 34 & 8  & 0  & 0  \\
        Suspiciousness&72  & 12 & 23 & 15 & 12 & 6  \\
        Hallucination&77  & 6  & 13 & 10 & 24 & 10 \\
        Unusual though content&74  & 11 & 8  & 20 & 24 & 3  \\
        Disorientation&116 & 19 & 5  & 0  & 0  & 0  \\
        Emotional withdrawal&51  & 52 & 28 & 9  & 0  & 0  \\
        Conceptual disorganization&65  & 32 & 23 & 9  & 11 & 0  \\
        Tension&82  & 40 & 16 & 2  & 0  & 0  \\
        Mannerism-Posturing&118 & 11 & 10 & 1  & 0  & 0  \\
        Motor retardation&104 & 18 & 17 & 1  & 0  & 0  \\
        Uncooperativeness&121 & 15 & 2  & 2  & 0  & 0  \\
        Blunted affect&85  & 19 & 28 & 8  & 0  & 0  \\
        Excitement&115 & 9  & 11 & 5  & 0  & 0 \\
    \bottomrule
    \end{tabular}
\end{table}

\section{Method}

\subsection{Data Preprocessing}

The audio recordings in the dataset were initially diarized to get the subject's speech alone from the recordings as the recordings also contained the interviewer's speech in them as well. After obtaining the subject's speech they were segmented into 40-second segments and the 40-second segments were then used for all speech and video based feature extraction purposes. In the context of text, all the interview sessions were transcribed manually through a third-party transcription service and the text corresponding to the subject's speech was only used for all text-based feature extraction for the models.

As for target values, due to the limited size of the dataset affecting the variability of the severity scores across different symptoms we moved to put all severity scores into 3 classes as shown in Table.\ref{tab:severity_scores}. This 3-class structure was selected based on the availability of severity scores in the dataset. As the dataset didn't include any hospitalized subjects, the dataset didn't contain any samples with predominantly extremely severe symptoms (a score of 7 on the BPRS scale). 

\begin{table}[h!]
  \caption{\centering
  \textbf{Symptom severity classes}}
  \label{tab:severity_scores}
  \centering
  \begin{tabular}{c c}
    \toprule
    BPRS score range & Class\\
    \midrule
    1&No symptoms\\
    2,3&Very mild \& mild symptoms \\
    4,5,6&Moderate \& severe symptoms\\
    \bottomrule
  \end{tabular}
\end{table}

\subsection{Speech-based feature extraction}

For speech based features, previous works on mental health assessment have shown that different features like pre-trained self-supervised acoustic features, articulatory coordination features and concise articulatory representations have shown promising performance in speech-based assessment across various tasks. So in this work, we extracted all these features and tested them out on different unimodal models. For pre-trained self-supervised acoustic features we extracted Wav2Vec2.0 \cite{wav2vec} and WavLM \cite{Wavlm} features. For articulatory features, we extracted vocal tract variables (TVs) from an acoustic to articulatory speech inversion system \cite{speechinversion} along with source articulatory features from an Aperiodicity, Periodicity, Pitch detector \cite{appdetector}. These TVs were then converted into Full Vocal Tract coordination (FVTC) features using a time-delay channel correlation mechanism \cite{FVTC}. Previous studies has been shown that changes in mental health conditions affect the coordination between articulatory gestures and these coordination changes are captured using these correlation features \cite{siriwardena2021multimodal}. In addition, since the time-delay correlation matrices are sparse and contain redundant information, we extracted a Concise Articulatory (C-Art) representations using a vector Quantized Variational Auto Encoder (VQ-VAE)-based representation learning model on the FVTC features \cite{gowtham_spade}. For session-level representations, the extracted features from all 40-second segments of the session were stacked together and padded to match the duration of the longest session in the dataset.

\begin{figure*}[th!]
  \centering
  \includegraphics[width=\linewidth]{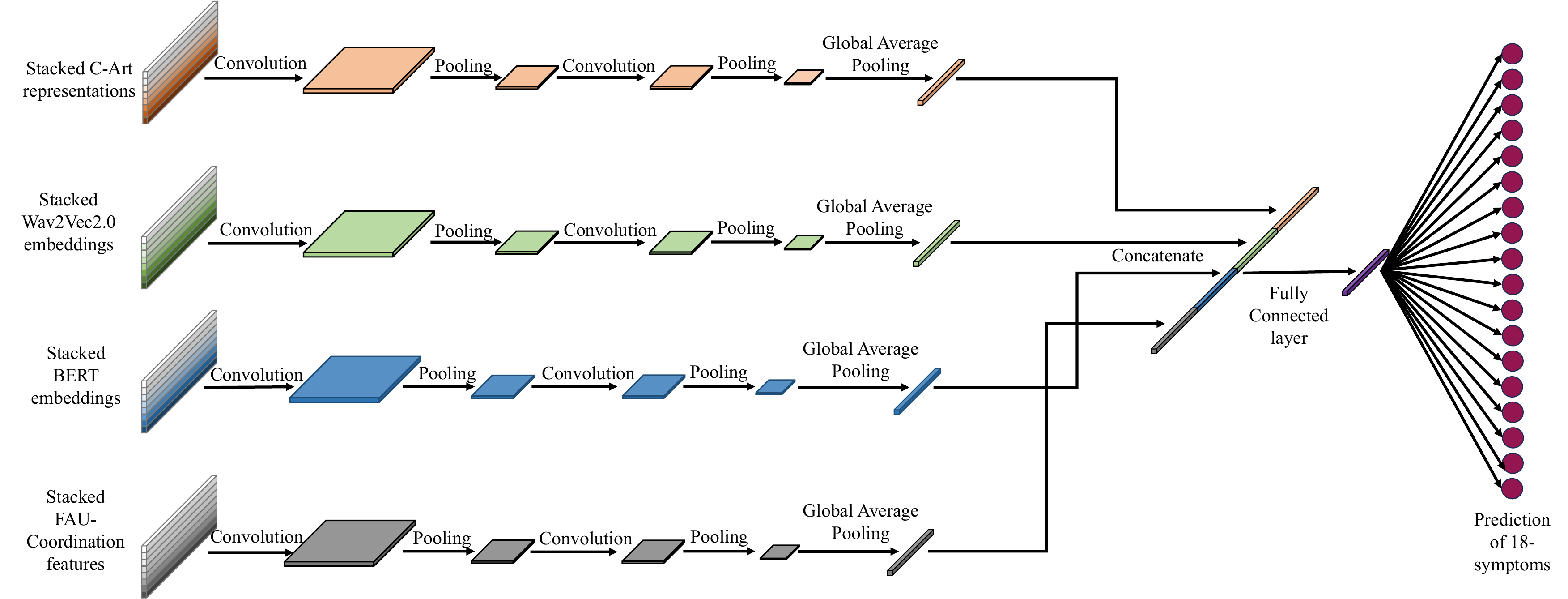}
  \caption{Multimodal model architecture with Wav2vec2.0, C-Art representations, BERT embeddings, and FAU-Coordination features.}
  \label{fig:multi}
\end{figure*}

\subsection{Video-based feature extraction}

For video-based features, we extracted Facial Action Units (FAU) \cite{prince2015facial} from Openface toolbox \cite{baltrusaitis2018openface}, the FAUs capture the movement of various facial muscles and facial landmarks. After extracting the FAUs, the FAU-based Coordination (FAUC) features were calculated similar to the way FVTC features were calculated from TVs using the same time-delay channel correlation mechanism \cite{FVTC}. To produce session-level representations, the extracted FAU-based coordination features from all 40-second segments of the session were stacked together and padded to match the duration of the longest session.

\subsection{Text-based feature extraction}

Text-based features were extracted from the transcripts of the subject's speech during the interview session using BERT encoder \cite{devlin2018bert}. The text was first split into sentences before processing. BERT embeddings were extracted using the BERT encoder, with the Class [CLS] token of each sentence used as the embedding for each sentence. To create session-level representations, BERT-based embeddings were stacked. Unlike the stacking of 40-second segments in audio and video features, for text embeddings of all the BERT-based embeddings corresponding to the sentences in the session were stacked together. To ensure consistency across sessions, all stacked representations were padded to match the length of the session with the highest sentence count.

\subsection{Unimodal models}

Unimodal models were independently trained and evaluated for speech, video, and text inputs. For speech-based analysis, multiple models were developed using pre-trained self-supervised acoustic features, FVTC features, and C-Art representations. Given previous findings \cite{gowtham_spade} that fusing pre-trained self-supervised acoustic features with C-Art representations for an Acoustic-Articulatory (AA)-Fusion approach improves performance on schizophrenia assessment task, we also trained and tested a similar AA-Fusion based speech model for this individual symptom severity estimation. For video and text inputs, unimodal models were trained using facial action unit (FAU)-based coordination features and BERT embeddings respectively. All the unimodal model architectures were designed with a Convolution Neural Network (CNN) backbone followed by a set of linear layers for multi-label multi-class predictions. The CNN style architecture was chosen because the stacked unimodal representations produced 2-dimensional inputs which can be effectively analyzed using the filters in CNN to capture spatial relationships and patterns among them.

\subsection{Multimodal models}

Multimodal models were trained based on the performance of the unimodal models. The best performing feature representations of speech, audio, and video were then fused together to create robust multimodal frameworks for individual schizophrenia symptom estimation. The model architecture of the best performing multimodal model is shown in Figure.\ref{fig:multi}. The multimodal model architecture was designed in a way that the best performing unimodal model architectures were used as branches without the final classification layer. The latent space representations from these branches were fused using a simple concatenation layer before being sent through a set of linear layers before making the final multi-label multi-class classification.

\begin{table*}[th!]
  \caption{\centering 
  \textbf{Summary of the classification models} (ACC: Accuracy, CI: Confidence Interval)}
  \label{tab:class_models}
  \centering
  \begin{tabular}{l l  c  c  c  c }
    \toprule
    {\textbf{Modality}} &{\textbf{Features}} & {\textbf{Fusion}}&\textbf{ACC with 95\% CI}&\textbf{Weighted F1 with 95\% CI} \\
    \midrule
    \multirow{6}{*}{Speech}&FVTC features&-&0.6228$\pm$0.0429&0.6211$\pm$0.0163\\
    &C-Art representations&-&0.6632$\pm$0.0495&0.6651$\pm$0.0415\\
    &Wav2Vec2.0&-&0.6344$\pm$0.0286&0.6363$\pm$0.0036\\
    &WavLM&-&0.6412$\pm$0.0662&0.6485$\pm$0.0483 \\
    &\textbf{C-Art with Wav2vec2.0} &\textbf{AA-fusion}&\textbf{0.6644$\pm$0.0621}&\textbf{0.6655$\pm$0.0398}\\
    &C-Art with WavLM &AA-fusion&0.6578$\pm$0.0707&0.6595$\pm$0.0.0509\\
    \midrule
    {Text}&BERT&-&0.6355$\pm$0.0224&0.6261$\pm$0.0143\\
    \midrule
    Video&FAU-Coordination &-&0.6227$\pm$0.0648&0.6108$\pm$0.0822\\
    \midrule
    \textbf{Multimodal}&\textbf{C-Art with Wav2vec2.0, BERT, FAU-coordination}&Late fusion&\textbf{0.7030$\pm$0.0495}&\textbf{0.7162$\pm$0.0283}\\
    \bottomrule
  \end{tabular}  
\end{table*}

\section{Experiments}

The dataset was divided into 3 equal folds and all the experiments were done and evaluated using a 3-fold cross validation where 2 folds were used for model development and the third being used for testing. The folds were hand crafted from the available sessions to make each fold has sessions with varying severity and sessions from both healthy controls and schizophrenia subjects available in the dataset.

Hyperparameter tuning for the unimodal and multimodal models were performed using grid search. The learning rate was chosen from the options \{5e-4, 1e-4, 5e-5\}, and the learning rate restart epoch for the scheduler was selected from \{25, 50, 75\}. For the model architecture, the kernel size for the convolutional layers was tested across \{2, 3, 4, 5, 6\}, and the pooling layer structure was selected from \{Average pooling , Max pooling\}.

From the results of the hyperparameter tuning, the best-performing multimodal model architecture included convolution layers with a kernel size of 5 and max pooling layer after the convolution layer. The model was trained for 200 epochs with an initial learning rate of 1e-4. The model was trained with a "Cosine Annealing with warm restarts" \cite{cosine} learning rate scheduler that had a restart set at every 50 epochs with the lowest possible learning rate set at 1e-6 and a learning rate multiplier after restart set at 2.

After model training all the models were evaluated based on symptom-wise accuracy, overall accuracy, and overall weighted F1-score. The symptom-wise scores of all the symptoms are not reported in the paper due to space constraints as each of the model will produce scores for each of the 18 symptoms. But they have been used for the in-depth analysis of the robustness of the models to predict severity of each symptom, and only the scores involved in these analysis are reported in this work.

\section{Results \& Discussion}

Table\ref{tab:class_models} shows the summary of the overall accuracy and overall weighted-F1 scores produced by the all the best-performing models for each feature representation and modalities discussed in this study. From the results in the table it is evident that Articulatory-Acoustic fusion approach between C-Art representations and Wav2vec2.0 embeddings perform better than all other speech representations across both evaluation metrics. And another important thing to notice from the results of all the unimodal models speech-based models fare better than the text and video based models in overall accuracy and overall weighted-F1 scores. But this might not be the case when looked at a granular level performance of these models on the symptoms individually.

We conducted an individual symptom analysis on the best-performing unimodal models to assess which symptoms each modality characterized most effectively. The analysis revealed that the "Hallucination" and "Unusual thought content" symptoms were more accurately captured by the unimodal audio model compared to both the text and video models. This indicates that speech-based features, such as vocal tone and patterns, are particularly suited for identifying these symptoms. In contrast, the "Emotional withdrawal" symptom was more accurately estimated by the top-performing video model, which suggests that facial expressions and non-verbal cues play a critical role in detecting this symptom. These findings underscore the importance of a multimodal approach, as each modality excels at capturing specific symptoms with greater accuracy, while others may be less effective.

The multimodal model, which combines speech, text, and video inputs, outperforms all unimodal models across both evaluation metrics, highlighting its superior capacity to capture a wider range of symptoms, despite using a relatively simple fusion mechanism. Furthermore, when analyzing symptom-wise performance, the multimodal model shows notable improvements across almost all individual symptoms compared to the best-performing unimodal model for those specific symptoms. Only two symptoms showed a slight performance degradation from their best-performing unimodal model, but the multimodal model still outperformed the other unimodal models, even in those cases. This reinforces the idea that integrating multiple modalities allows for a more comprehensive and accurate assessment of schizophrenia symptoms, improving diagnostic accuracy and providing more reliable insights for clinicians.   

\section{Conclusion \& Future work}

This study focused on estimating the severity of individual schizophrenia symptoms using unimodal and multimodal machine learning models. By incorporating speech, video, and text-based inputs, we aimed to create a scalable framework for individual symptom assessment, moving beyond traditional subjective clinical evaluations and deep-learning based binary classification approaches.

Our experiments demonstrated that unimodal models, while effective in capturing symptom-related features, varied in performance depending on the feature representations used. And based on the symptom analysis, it was also noted that certain modalities captured some symptoms better than others. This result stood as proof that individual modalities had limitations in fully capturing the complexity of schizophrenia symptoms. To address this, we developed a multimodal model that integrated the strongest unimodal models with different feature representations. The results showed that this approach improved accuracy and robustness in symptom severity estimation, reinforcing the idea that combining multiple modalities provides a more comprehensive understanding of schizophrenia symptoms.

While our study demonstrates the potential of multimodal learning in schizophrenia assessment, there are several areas for further improvement. Future work could explore advanced fusion techniques beyond simple concatenation to optimize feature integration. Expanding the dataset with more diverse and larger samples would enhance model generalizability. Finally, clinical validation is essential to ensure the model’s practical applicability in real-world mental health settings.

\section{Acknowledgment}

This work was supported by the National Science Foundation grant numbered 2124270.

\bibliographystyle{IEEEtran}
\bibliography{mybib}

\end{document}